\documentclass[12pt]{iopart}

\expandafter\let\csname equation*\endcsname\relax
\expandafter\let\csname endequation*\endcsname\relax
\usepackage{amsmath}
\usepackage{bm}

\begin{document}

\title{Scattering series in mobility problem for suspensions}

\author{Karol Makuch}

\address{Institute of Theoretical Physics, Faculty of Physics, University of Warsaw,
Ho\.{z}a 69, 00-681 Warsaw, Poland}

\ead{Karol.Makuch@fuw.edu.pl}

\begin{abstract}
The mobility problem for suspension of spherical particles immersed in an
arbitrary flow of a viscous, incompressible fluid is considered in the regime
of low Reynolds numbers. The scattering series which appears in the mobility
problem is simplified. The simplification relies on the reduction of the
number of types of single-particle scattering operators appearing in the
scattering series. In our formulation there is only one type of
single-particle scattering operator.

\end{abstract}

\maketitle

\section{Introduction}

Various types of suspensions of spherical particles can be found in nature and
industry. The diversity follows from the facts that systems may be composed of
many sorts of particles, such as hard-spheres or spherical polymers
\cite{fernandez2010microgel}, and many types of forces may act between
particles \cite{russel1992colloidal}, \cite{Odenbach2009Colloidal}. This
variety in a structure implies a wide range of phenomena exhibited by
suspensions. To understand them it is often crucial to consider hydrodynamic
interactions between particles, that is their mutual influence through
movement of the surrounding liquid (which is different than the influence
through direct forces, such as magnetic or van der Waals force). In many
physical situations examination of the hydrodynamic interactions amounts to
the friction problem or the mobility problem \cite{kim1991microhydrodynamics}.
In both cases the particles are immersed in a flow of the surrounding liquid
(ambient flow). In the friction problem the velocities of particles are
assumed and the hydrodynamic forces on the fluid produced by the particles are
calculated. This problem appears e.g. in the determination of Stokes
coefficient for polymer modeled as an agglomerate of spherical particles
\cite{debye1948intrinsic, brinkman1949calculation, felderhof1975frictional}.
In the mobility problem one determines the velocities of freely moving
particles and also the hydrodynamic forces acting on the fluid. The particles
are assumed to be immersed in the ambient flow and the external forces may act
on them. Among the situations in which the mobility problem appears, the
determination of sedimentation coefficient of the suspension can be mentioned
\cite{batchelor1972sedimentation, batchelor1976brownian}.

To solve both the friction and the mobility problem one starts with the
equations which govern the dynamics of the suspension. Here we assume linear,
stationary Stokes equations for an incompressible fluid with the stick
boundary conditions on the surface of particles. One of the possible
approaches to the Stokes equations is the method of successive approximations
\cite{lamb1997hydrodynamics}, also known as the reflection method, developed
by Smoluchowski \cite{Smoluchowski1912practical}. It is based on linearity of
the Stokes equations. The method consists in successive superpositions of the
single-particle solutions of the Stokes equations chosen to fulfill the
boundary conditions with the increasing accuracy. The above procedure leads to
the solution of the friction problem in a form of the superpositions of
multiple reflected flows. The resulting structure of the solution is called the scattering series. In
the series the single-particle operator plays an important role. In case of
the friction problem the scattering series has a simple form because there is
only one type of the single-particle operator. The situation is more
complicated in the mobility problem. Here four types of single-particle
operators are considered \cite{diag_ostateczna_wersja}. Due to the difference
in the number of single-particle operators statistical physics considerations
are easier in the case of the friction problem than in the case of the
mobility problem. For the latter, many formulas of the same structure can be
found in the literature \cite{cichocki2008stokesian, szymczak2004memory,
mazur1982many, felderhof1988sedimentation, diag_ostateczna_wersja,
resummation83, Beenakker1984effective, Beenakker1984Diffusion}.

The aim of the present article is the reformulation of the mobility problem.
The reformulation results in the scattering series in which only one
type of single-particle operator appears. It enables simplification of the statistical
physics considerations relevant to the mobility problem.

\section{Governing equations}

The system under consideration consists of $N$ identical hard spherical
particles of radius $a$ immersed in an incompressible, infinite Newtonian
liquid with kinematic viscosity $\eta$. The inertia of particles and the
inertia terms in the incompressible Navier-Stokes equations are assumed to be
negligible. As a consequence, the fluid is governed by the steady Stokes
equations \cite{kim1991microhydrodynamics}. We supplement the Stokes equations
with the stick (no-slip) boundary conditions on the surface of particles
\cite{lauga_brenner_stone}. Next, following idea of Mazur and Bedeaux
\cite{Mazur1974235} we extend the Stokes equations inside the particles in the
following way:
\begin{subequations}
\label{rownania stokesa}%
\begin{align}
\bm{\nabla}p\left(  \bm{r}\right)  -\eta\Delta\bm{v}\left(
\bm{r}\right)   &  =\bm{f}\left(  \bm{r}\right)
,\label{jednorodne rownania stokesa 1}\\
\bm{\nabla}\cdot\bm{v}\left(  \bm{r}\right)   &  =0,
\label{jednorodne rownania stokesa 2}%
\end{align}
introducing induced force densities $\bm{f}\left(  \bm{r}\right)  $
\cite{cox1967effect}, \cite{cox1971rheology}. Here $p\left(  \bm{r}%
\right)  $ and $\bm{v}\left(  \bm{r}\right)  $ are the pressure field and
the velocity field of the suspension. The induced force densities are
determined \cite{felderhof1976force} by the condition that the flow of the
suspension inside the particles reproduces their hard-sphere velocity field,%
\end{subequations}
\begin{equation}
\bm{v}\left(  \bm{r}\right)  =\bm{U}_{i}\left(  \bm{r}\right)
=\bm{V}_{i}+\bm{\Omega}_{i}\times\left(  \bm{r}-\bm{R}%
_{i}\right)  \ \ \ \ \ \ \ \ \ \text{for }\left\vert \bm{r}-\bm{R}%
_{i}\right\vert \leq a, \label{def U field}%
\end{equation}
where $\bm{V}_{i}$ and $\bm{\Omega}_{i}$ is translational and
rotational velocity of the $i$-th particle which is located at the position
$\bm{R}_{i}$.

For the case under consideration the force densities $\bm{f}\left(
\bm{r}\right)  $ are localized only on the surface of particles
\cite{mazur1982many}, \cite{marysiaElekFragment}, that is%
\begin{equation}
\bm{f}\left(  \bm{r}\right)  =\sum_{i=1}^{N}\bm{f}_{i}\left(
\bm{r}\right)  , \label{force f by fi}%
\end{equation}
with $\bm{f}_{i}\left(  \bm{r}\right)  $ localized only on the surface
of the $i$-th particle,%
\begin{equation}
\bm{f}_{i}\left(  \bm{r}\right)  =-\bm{\sigma}\cdot\bm{n}%
_{i}\bm{\ }\delta\left(  \left\vert \bm{r-R}_{i}\right\vert -a\right)
, \label{sila indukowana a tensor cisnien}%
\end{equation}
where $\bm{\sigma}$ is the stress tensor for the fluid
\cite{kim1991microhydrodynamics}, $\bm{n}_{i}$ - a vector normal to the
surface of the particle $i$ at point $\bm{r}$, and $\delta\left(
x\right)  $ - Dirac delta function.

The above extension of the Stokes equations allows to use the Green function
method. The method leads to the expression for the flow of suspension in the
whole space \cite{ladyzhenskaya:57},%
\begin{equation}
\bm{v}\left(  \bm{r}\right)  =\int d^{3}\bm{r}^{\prime}%
\bm{G}_{0}\left(  \bm{r}-\bm{r}^{\prime}\right)  \bm{\cdot
f}\left(  \bm{r}^{\prime}\right)  , \label{rozwiazanie formalne}%
\end{equation}
where Oseen tensor $\bm{G}_{0}$ has the following form
\cite{pozrikidis1992boundary}%

\begin{equation}
\bm{G}_{0}\left(  \bm{r}\right)  =\frac{1}{8\pi\eta}\frac
{\bm{1}+\bm{\hat{r}\hat{r}}}{\left\vert \bm{r}\right\vert
},\hspace{1in}\bm{\hat{r}}=\frac{\bm{r}}{\left\vert \bm{r}%
\right\vert }. \label{tensor Oseena}%
\end{equation}

\section{Friction problem}

In the friction problem the particles are immersed in a liquid in which
initially ambient flow $\bm{v}_{0}\left(  \bm{r}\right)  $ is present
and their translational $\bm{V}_{i}$ and rotational velocities
$\bm{\Omega}_{i}$ are assumed to be known. The aim of the friction problem
is a determination of the force densities $\bm{f}_{i}\left(
\bm{r}\right)  $ induced on the surface of particles.

For a single particle, the friction problem for a particular type of ambient
flow was solved more than a century ago \cite{lamb1997hydrodynamics}. It was
later generalized for an arbitrary ambient flow \cite{felderhof1976force},
\cite{schmitz1982creeping}. The solution of the Stokes equations
(\ref{rownania stokesa}) in this case has a form of the following linear
relation \cite{schmitz1982creeping}%
\begin{equation}
\bm{f}_{1}\left(  \bm{r}\right)  =\int d^{3}\bm{r}^{\prime
}\bm{Z}_{0}\left(  \bm{R}_{1},\bm{r},\bm{r}^{\prime}\right)
\bm{\cdot}\left(  \bm{U}_{1}\left(  \bm{r}^{\prime}\right)
-\bm{v}_{0}\left(  \bm{r}^{\prime}\right)  \right)
.\label{opor jedna czastka calkowa}%
\end{equation}
Single particle resistance operator $\bm{Z}_{0}$ is localized on the
surface of a particle. It means that the force density $\bm{f}_{1}\left(
\bm{r}\right)  $ is localized on the particle surface - consistently with the
equation (\ref{sila indukowana a tensor cisnien}) - and its value depends only
on the velocity field $\bm{U}_{1}\left(  \bm{r}\right)  -\bm{v}%
_{0}\left(  \bm{r}\right)  $ at points $\left\vert \bm{r}%
-\bm{R}_{1}\right\vert =a$. The details of the resistance operator
$\bm{Z}_{0}$ can be found in the Appendix. The equation
(\ref{opor jedna czastka calkowa}) will have the following form in shorthand notation%

\begin{equation}
\bm{f}_{1}=\bm{Z}_{0}\left(  1\right)  \bm{\cdot}\left(
\bm{U}_{1}-\bm{v}_{0}\right)  , \label{opor jedna czastka}%
\end{equation}
in which the integral variables are omitted and the position of a particle is
denoted by its index. The notation will be extensively used further on.

It is worth mentioning that the single particle friction problem for a
spherical shape has been solved not only for a hard sphere with the stick
boundary conditions. Many other physical situations have also been considered
in the literature, e.g. different boundary conditions
\cite{schmitz1978creeping}, permeable particles \cite{felderhof1978faxen},
spherical polymers \cite{felderhof1990hydrodynamics}, immiscible droplets
\cite{geigenmuller1986many}, \cite{cichocki1988hydrodynamic} or more complex
cases \cite{cichocki2009hydrodynamic}, \cite{blawzdziewicz1999hydrodynamic}.
In general, the relation (\ref{opor jedna czastka}) is still valid but with
modified $\bm{Z}_{0}$ operator.

The concept of the induced force densities and linearity of the Stokes
equations allows to use the single particle friction problem to find the
solution of the friction problem for suspension. In fact, the $i$-th particle
in suspension is surrounded by a flow which is a superposition of the ambient
flow $\bm{v}_{0}\left(  \bm{r}\right)  $ and the flow induced by other
particles, $\sum_{j\not =i}\int d\bm{r}^{\prime}\ \bm{G}_{0}\left(
\bm{r},\bm{r}^{\prime}\right)  \bm{f}_{j}\left(  \bm{r}%
^{\prime}\right)  $. Applying this modified ambient flow to equation
(\ref{opor jedna czastka}) leads to the following expression%
\begin{equation}
\bm{f}_{i}=\bm{Z}_{0}\left(  i\right)  \bm{\cdot}\left(
\bm{U}_{i}-\bm{v}_{0}-\sum_{j\not =i}\bm{G}_{0}\bm{f}%
_{j}\right)  \label{opor zawiesina}%
\end{equation}
written in the shorthand notation. This is the formula where one can directly
implement the reflection method by successive iterations. It yields%
\begin{equation}
\bm{f}_{i}=\sum_{j=1}^{N}\bm{Z}_{ij}\left(  1\ldots N\right)  \left(
\bm{U}_{j}-\bm{v}_{0}\right)  , \label{solution of fric prob}%
\end{equation}
where the friction operator $\bm{Z}_{ij}$ has the form of the scattering
series%
\begin{eqnarray}
\bm{Z}_{ij}\left(  1\ldots N\right)  &=&\delta_{ij}\bm{Z}_{0}\left(
i\right)  -\left(  1-\delta_{ij}\right)  \bm{Z}_{0}\left(  i\right)
\bm{G}_{0}\left(  ij\right)  \bm{Z}_{0}\left(  j\right)  + \nonumber \\
&&+\sum
_{k}^{\prime}\bm{Z}_{0}\left(  i\right)  \bm{G}_{0}\left(  ik\right)
\bm{Z}_{0}\left(  k\right)  \bm{G}_{0}\left(  kj\right)
\bm{Z}_{0}\left(  j\right)  +\ldots\label{Z scattering series}%
\end{eqnarray}
The different terms in the equation (\ref{Z scattering series}) correspond to
scattering sequences. The prime symbol indicates summation over $k$ different
than neighboring particle indexes in the scattering sequence.

\section{Mobility problem}

In the mobility problem one considers freely moving particles immersed in an
ambient flow of the fluid $\bm{v}_{0}\left(  \bm{r}\right)  $ and
subjected to the action of the external forces. The aim of the problem is to
calculate the velocity fields of the particles $\bm{U}_{i}\left(
\bm{r}\right)  $ and also the induced force densities $\bm{f}%
_{i}\left(  \bm{r}\right)  $ \cite{kim1991microhydrodynamics}.

In order to obtain the solution for a suspension we first analyze a case of a
single particle. Before going into the details, it should be noticed that
linearity of the Stokes equations implies linear relation between the response
of the particle and the source of a disturbance. Therefore the velocity field
of the particle $\bm{U}_{i}\left(  \bm{r}\right)  $ and the induced
forces $\bm{f}_{i}\left(  \bm{r}\right)  $ on its surface are linear
to the ambient flow $\bm{v}_{0}\left(  \bm{r}\right)  $ and to the
external forces $\bm{f}_{ext}\left(  \bm{r}\right)  $,%
\begin{equation}
\bm{U}_{1}\left(  \bm{r}\right)  =\int d\bm{r}^{\prime}%
\bm{M}_{0}\left(  \bm{R}_{1}\bm{,r,r}^{\prime}\right)
\bm{f}_{ext}\left(  \bm{r}^{\prime}\right)  +\int d\bm{r}^{\prime
}\bm{M}_{<}\left(  \bm{R}_{1}\bm{,r,r}^{\prime}\right)
\bm{v}_{0}\left(  \bm{r}^{\prime}\right)  , \label{U1 single}%
\end{equation}%
\begin{equation}
\bm{f}_{1}\left(  \bm{r}\right)  =\int d\bm{r}^{\prime}%
\bm{M}_{>}\left(  \bm{R}_{1}\bm{,r,r}^{\prime}\right)
\bm{f}_{ext}\left(  \bm{r}^{\prime}\right)  +\int d\bm{r}^{\prime
}\bm{\hat{M}}\left(  \bm{R}_{1}\bm{,r,r}^{\prime}\right)
\bm{v}_{0}\left(  \bm{r}^{\prime}\right)  . \label{f1 single}%
\end{equation}
Since the above statement is based only on the linearity of the governing
equations it is correct for different particles and boundary conditions
mentioned in the previous section. The single particle operators
$\bm{M}_{0}$, $\bm{M}_{<}$, $\bm{M}_{>}$ and $\bm{\hat{M}}$
need to be determined for every particular case. In what follows we discuss
the hard spheres with the stick boundary conditions
\cite{felderhof1988sedimentation}, \cite{diag_ostateczna_wersja}. From the
papers cited here, one can easily infer the form of $\bm{M}_{0}%
\equiv\bm{\mu}_{0}$ operator which is explicitly given in the Appendix.
The remaining operators are expressed with the following equations%
\begin{align}
\bm{M}_{<}\left(  1\right)   &  =\bm{\mu}_{0}\left(  1\right)
\bm{Z}_{0}\left(  1\right)  ,\\
\bm{M}_{>}\left(  1\right)   &  =\bm{Z}_{0}\left(  1\right)
\bm{\mu}_{0}\left(  1\right)  ,\\
\bm{\hat{M}}\left(  1\right)   &  =-\bm{Z}_{0}\left(  1\right)
+\bm{Z}_{0}\left(  1\right)  \bm{\mu}_{0}\left(  1\right)
\bm{Z}_{0}\left(  1\right)
\end{align}
which were written in the shorthand notation.

Due to linearity of the Stokes equations the above single particle solution of
the mobility problem can be used to solve the case of suspension. In the
suspension, the $i$-th particles is immersed in the flow given by a
superposition of the ambient flow $\bm{v}_{0}\left(  \bm{r}\right)  $
and the flow generated by other particles, $\sum_{j\not =i}\int d\bm{r}%
^{\prime}\ \bm{G}_{0}\left(  \bm{r},\bm{r}^{\prime}\right)
\bm{f}_{j}\left(  \bm{r}^{\prime}\right)  $. Assuming this modified
ambient flow in equations (\ref{U1 single}) and (\ref{f1 single}) leads to the
expressions for the velocity of particles%
\begin{equation}
\bm{U}_{i}=\bm{M}_{0}\left(  i\right)  \bm{f}_{ext}+\bm{M}%
_{<}\left(  i\right)  \left(  \bm{v}_{0}+\sum_{j\not =i}\bm{G}%
_{0}\bm{f}_{j}\right)  \label{ui mobility problem}%
\end{equation}
and the induced force densities in suspension%
\begin{equation}
\bm{f}_{i}=\bm{M}_{>}\left(  i\right)  \bm{f}_{ext}+\bm{\hat
{M}}\left(  i\right)  \left(  \bm{v}_{0}+\sum_{j\not =i}\bm{G}%
_{0}\bm{f}_{j}\right)  . \label{Fi mobility problem}%
\end{equation}

In what follows we rewrite the equations (\ref{ui mobility problem}) and
(\ref{Fi mobility problem}) in the following concise form%

\begin{equation}
\bm{s}_{i}=\bm{M}\left(  i\right)  \left(  \bm{\psi}_{0}%
+\sum_{j\not =i}\bm{Gs}_{j}\right)  , \label{s uklad}%
\end{equation}
where the response of the $i$-th particle $\bm{s}_{i}$, single-particle
mobility operator $\bm{M}\left(  i\right)  $, and external field
$\bm{\psi}_{0}$ are defined respectively below:%
\begin{equation}
\bm{s}_{i}=%
\begin{bmatrix}
\bm{U}_{i}\\
\bm{f}_{i}%
\end{bmatrix}
, \label{def S}%
\end{equation}%
\begin{equation}
\bm{\psi}_{0}=%
\begin{bmatrix}
\bm{F}_{ext}\\
\bm{v}_{0}%
\end{bmatrix}
, \label{def psi0}%
\end{equation}%
\begin{equation}
\bm{M}\left(  i\right)  =%
\begin{bmatrix}
\bm{M}_{0}\left(  i\right)  & \bm{M}_{<}\left(  i\right) \\
\bm{M}_{>}\left(  i\right)  & \bm{\hat{M}}\left(  i\right)
\end{bmatrix}
, \label{def M}%
\end{equation}
whereas $\bm{G}$ is generalized Oseen tensor $\bm{G}_{0}$:%
\begin{equation}
\bm{G}=%
\begin{bmatrix}
0 & 0\\
0 & \bm{G}_{0}%
\end{bmatrix}
. \label{def G}%
\end{equation}

The method of iterations applied to the equations (\ref{s uklad}) leads to the
following solution of the mobility problem%

\begin{equation}
\bm{s}_{i}\left(  \bm{R}_{1},\ldots,\bm{R}_{N}\right)  =\sum
_{j=1}^{N}\bm{T}_{ij}\left(  \bm{R}_{1},\ldots,\bm{R}_{N}\right)
\bm{\psi}_{0},
\end{equation}
where $\bm{T}_{ij}$ is given by the scattering series as follows%
\begin{eqnarray}
\bm{T}_{ij}\left(  \bm{R}_{1},\ldots,\bm{R}_{N}\right)
&=&\delta_{ij}\bm{M}\left(  \bm{R}_{i}\right)  +\left(  1-\delta
_{ij}\right)  \bm{M}\left(  \bm{R}_{i}\right)  \bm{GM}\left(
\bm{R}_{j}\right)  + \nonumber \\
&&+\sum_{k}^{\prime}\bm{M}\left(  \bm{R}%
_{i}\right)  \bm{GM}\left(  \bm{R}_{k}\right)  \bm{GM}\left(
\bm{R}_{j}\right)  +\ldots\label{reformulated ss}%
\end{eqnarray}
It is worth comparing the scattering series given by expression
(\ref{reformulated ss}) to the scattering series%
\begin{equation}
\delta_{ij}\bm{M}_{0}\left(  \bm{R}_{i}\right)  +\left(  1-\delta
_{ij}\right)  \bm{M}_{<}\left(  \bm{R}_{i}\right)  \bm{GM}%
_{>}\left(  \bm{R}_{j}\right)  +\sum_{k}^{\prime}\bm{M}_{<}\left(
\bm{R}_{i}\right)  \bm{G\hat{M}}\left(  \bm{R}_{k}\right)
\bm{GM}_{>}\left(  \bm{R}_{j}\right)  +\ldots
\label{mobility problem scattering}%
\end{equation}
which is considered in the literature \cite{diag_ostateczna_wersja}. Notice
that here four types of single-particle operators $\bm{M}_{0}$,
$\bm{M}_{<}$, $\bm{M}_{>}$, and $\bm{\hat{M}}$ appear. With
respect to the number of types of single-particle operators, the formulation
of the scattering series (\ref{reformulated ss}) introduced in the present
paper is simpler than the series given by expression
(\ref{mobility problem scattering}).

It is worth mentioning that no approximation was made in the above analysis.
In particular, the equation (\ref{s uklad}) is a proper starting point
to analyse hydrodynamic interactions of particles in close contact.

\section{Discussion}

In the present paper the scattering series for the mobility problem has been
reformulated which results in the simple form given by the equation
(\ref{reformulated ss}). The simplification relies on the fact that in
expression (\ref{reformulated ss}) there is only one type of single-particle
operator, $\bm{M}$. In the formulation hitherto used in the literature
\cite{diag_ostateczna_wersja}, in the scattering series there are four type of
single-particle operators $\bm{M}_{0}$, $\bm{M}_{<}$, $\bm{M}_{>}%
$, and $\bm{\hat{M}}$ which is showed by the expression
(\ref{mobility problem scattering}).

At first sight the difference between both formulations may not seem to be
significant. However, the mobility problem plays a crucial role e.g. for
calculations of transport coefficients of suspensions. In this context
statistical physics considerations contain many different formulas of the same
structure \cite{felderhof1988sedimentation}, \cite{diag_ostateczna_wersja},
\cite{cichocki2008stokesian}, \cite{szymczak2004memory}. A forcible example of
cumbersomeness following from a lack of a simple formulation of the mobility
problem is the Beenakker-Mazur method \cite{resummation83},
\cite{Beenakker1984effective}, \cite{Beenakker1984Diffusion} which is used to
calculate short-time dynamic properties of suspensions. A multitude of
expressions occurring in cited articles obscures the essence of the method
which, on the other hand, is the most comprehensive analytical scheme
available so far \cite{gapinski2009structure}. With the aim of the
reformulated scattering sequence the Beenakker-Mazur method$\ $can be
presented in a simple form which we are going to show in another paper. The
simple formulation of the mobility problem also allows to carry out advanced
analysis of scattering series in a clear and simple way. In this context the
formulation will also be used in our subsequent work on macroscopic characteristics
of suspensions.

\section*{Acknowledgements}

The author thanks Krzysztof Byczuk for his suggestions. The author also acknowledges support by the Foundation for Polish Science (FNP)
through the TEAM/2010-6/2 project co-financed by the EU European Regional
Development Fund.%

\appendix

\section{Single particle operators and multipole picture}

Here we give the explicit form of $\bm{Z}_{0}$ and $\bm{\mu}_{0}$
operators. Basing on the reference \cite{marysiaElekFragment} the
$\bm{Z}_{0}$ operator can be expressed by the following formula
\begin{eqnarray}
 \bm{Z}_{0}\left(  \bm{R,r},\bm{r}^{\prime}\right) =& \sum
_{l,l^{\prime}=1}^{\infty}\sum_{m^{\prime}=-l^{\prime}}^{l^{\prime}}%
\sum_{m=-l}^{l}\sum_{\sigma,\sigma^{\prime}=0}^{2} \delta_{a}\left(
\bm{r-R}\right)  \bm{w}_{lm\sigma}^{+}\left(  \bm{r-R}\right) \nonumber \\
&\times \left[  Z_{0}\left(  \bm{R}\right)  \right]_{lm\sigma,l^{\prime
}m^{\prime}\sigma^{\prime}}
 \delta_{a}\left(  \bm{r}^{\prime}%
\bm{-R}\right)  \bm{w}_{l^{\prime}m^{\prime}\sigma^{\prime}}^{+\ast
}\left(  \bm{r}^{\prime}\bm{-R}\right)  ,  \label{Z0 expression}%
\end{eqnarray}
where $\left[  Z_{0}\left(  \bm{R}\right)  \right]  _{lm\sigma,l^{\prime
}m^{\prime}\sigma^{\prime}}$ stands for the multipole matrix
with indexes $l=1,\ldots,\infty;$ $m=-l,\ldots,l;$
$\sigma=0,1,2$. 
Its matrix elements are explicitly given e.g. in the reference \cite{Cichocki2002three}.
A set of multipole functions $\bm{v}_{lm\sigma}^{+}\left(  \bm{r}\right)  $ and
$\bm{w}_{lm\sigma}^{+}\left(  \bm{r}\right)  $
is defined e.g. in the references \cite{cichocki2000friction} or \cite{marysiaElekFragment}.
Every
solution of the homogeneous Stokes equations may be expressed as a
combinations of the multipole functions $\bm{v}_{lm\sigma}^{+}\left(
\bm{r}\right)  $%
\begin{equation}
\bm{v}_{0}\left(  \bm{r}\right)  =\sum_{l=1}^{\infty}\sum_{m=-l}%
^{l}\sum_{\sigma=0}^{2}\left[  v_{0}\left(  \bm{R}\right)  \right]
_{lm\sigma}\bm{v}_{lm\sigma}^{+}\left(  \bm{r-R}\right)  ,
\label{v0 na multipole}%
\end{equation}
whereas the induced surface force $\bm{f}_{i}\left(  \bm{r}\right)  $
as combination of multipole functions $\bm{w}_{lm\sigma}^{+}\left(
\bm{r}\right)  $:%
\begin{equation}
\bm{f}_{i}\left(  \bm{r}\right)  =\sum_{l=1}^{\infty}\sum_{m=-l}%
^{l}\sum_{\sigma=0}^{2}\left[  f_{i}\right]  _{lm\sigma}\delta_{a}\left(
\bm{r}-\bm{R}_{i}\right)  \bm{w}_{lm\sigma}^{+}\left(
\bm{r}-\bm{R}_{i}\right)  . \label{f1 na multipole}%
\end{equation}
The multipole functions $\bm{w}_{lm\sigma}^{+}\left(
\bm{r}\right)  $ are defined as orthonormal to $\bm{v}_{lm\sigma}%
^{+}\left(  \bm{r}\right)  $ functions. This orthonormality is expressed
in the following way%
\begin{equation}
\left\langle \delta_{a}\bm{w}_{lm\sigma}^{+}|\bm{v}_{l^{\prime
}m^{\prime}\sigma^{\prime}}^{+}\right\rangle =\delta_{ll^{\prime}}%
\delta_{mm^{\prime}}\delta_{\sigma\sigma^{\prime}}, \label{ortonormalnosc}%
\end{equation}
with the Dirac notation \cite{cohen-quantum} for scalar product of two vector
fields $\bm{A}\left(  \bm{r}\right)  $ and $\bm{B}\left(
\bm{r}\right)  $:%
\begin{equation}
\left\langle \bm{A}|\bm{B}\right\rangle =\int d^{3}\bm{rA}^{\ast
}\left(  \bm{r}\right)  \cdot\bm{B}\left(  \bm{r}\right)  ,
\end{equation}
and the scalar function $\delta_{a}\left(  \bm{r}\right)  $ of the form%
\[
\delta_{a}\left(  \bm{r}\right)  \bm{=}a^{-1}\delta\left(  \left\vert
\bm{r}\right\vert -a\right)
\]
which confines integration area to the sphere of the radius $a$.

Operator $\bm{\mu}_{0}$ is given by the expression%
\begin{eqnarray}
\bm{\mu}_{0}\left(  \bm{R,r},\bm{r}^{\prime}\right)
=& \sum_{l,l^{\prime}=1}^{\infty}\sum_{m^{\prime}= -l^{\prime}}^{l^{\prime}}%
\sum_{m=-l}^{l}\sum_{\sigma,\sigma^{\prime}=0}^{2}\Theta_{a}\left(
\bm{r-R}\right)  \bm{v}_{lm\sigma}^{+}\left(  \bm{r-R}\right) \nonumber \\
& \times \left[  \mu_{0}\left(  \bm{R}\right)  \right]  _{lm\sigma,l^{\prime
}m^{\prime}\sigma^{\prime}}\Theta_{a}\left(  \bm{r}^{\prime}%
\bm{-R}\right)  \bm{v}_{l^{\prime}m^{\prime}\sigma^{\prime}}^{+\ast
}\left(  \bm{r}^{\prime}\bm{-R}\right)  \label{mu0 expression}%
\end{eqnarray}
with multipole matrix $\left[  \mu_{0}\left(  \bm{R}\right)  \right]
_{lm\sigma,l^{\prime}m^{\prime}\sigma^{\prime}}$ explicitly given in the
reference \cite{Cichocki2002three}. Here $\Theta_{a}\left(  \bm{r}%
-\bm{R}\right)  $ is a characteristic function of the particle at position
$\bm{R}$: it equals $0$ whenever $\bm{r}$ points outside the particle,
and equals $1$ otherwise.

Finally, we represent the equations (\ref{ui mobility problem}) and
(\ref{Fi mobility problem}) in the multipole expansion formalism. To pass on
to the multipole picture we put the expressions (\ref{Z0 expression}) and
(\ref{mu0 expression}) into these equations and multiply them by $\left\langle
\bm{w}_{lm\sigma}^{+}\left(  i\right)  \delta_{a}\left(  i\right)
\right\vert $ from the left side. A simple algebra yields
\begin{subequations}
\label{upr rozwiazania mult}%
\begin{align}
U_{i}  &  =\mu_{0}\left(  i\right)  f_{ext}\left(  i\right)  +\mu_{0}\left(
i\right)  Z_{0}\left(  i\right)  \left(  v_{0}\left(  i\right)  +\sum
_{j\not =i}G_{0}\left(  ij\right)  f_{j}\right)  ,\\
f_{i}  &  =Z_{0}\left(  i\right)  \mu_{0}\left(  i\right)  f_{ext}\left(
i\right)  -\hat{Z}_{0}\left(  i\right)  \left(  v_{0}\left(  i\right)
+\sum_{j\not =i}G_{0}\left(  ij\right)  f_{j}\right)  ,
\end{align}
where the multipole vector of the ambient flow at point $\bm{R}$%
\end{subequations}
\begin{equation}
\left[  v_{0}\left(  \bm{R}\right)  \right]  _{lm\sigma}=\left\langle
\bm{w}_{lm\sigma}^{+}\left(  \bm{R}\right)  \delta_{a}\left(
\bm{R}\right)  |\bm{v}_{0}\right\rangle , \label{def multipoli v}%
\end{equation}
multipole velocity field $U_{i}$ for the $i$-th particle%
\begin{equation}
\left[  U_{i}\right]  _{lm\sigma}=\left\langle \bm{w}_{lm\sigma}%
^{+}\left(  i\right)  \delta_{a}\left(  i\right)  |\bm{U}_{i}\right\rangle
, \label{def U}%
\end{equation}
induced surface force multipole $f_{i}$ for the $i$-th particle%
\begin{equation}
\left[  f_{i}\right]  _{lm\sigma}=\left\langle \bm{v}_{lm\sigma}%
^{+}\left(  i\right)  |\bm{F}_{i}\right\rangle , \label{def multipoli f}%
\end{equation}
and external force multipole field $f_{ext}\left(  \bm{R}\right)  $ at
point $\bm{R}$%
\begin{equation}
\left[  f_{ext}\left(  \bm{R}\right)  \right]  _{lm\sigma}=\left\langle
\bm{v}_{lm\sigma}^{+}\left(  \bm{R}\right)  \Theta_{a}\left(
\bm{R}\right)  |\bm{F}_{ext}\right\rangle .
\end{equation}
In the above formulas $\left\vert \bm{A}\left(  i\right)  \right\rangle $
or $\left\vert \bm{A}\left(  \bm{R}\right)  \right\rangle $ denote
vector fields $\bm{A}\left(  \bm{r-R}_{i}\right)  $ or $\bm{A}%
\left(  \bm{r-R}\right)  $ respectively. Moreover matrix $G_{0}\left(
\bm{R,R}^{\prime}\right)  $ is defined with the formula%
\begin{equation}
\left[  G_{0}\left(  \bm{R},\bm{R}^{\prime}\right)  \right]
_{lm\sigma,l^{\prime}m^{\prime}\sigma^{\prime}}=\left\langle \bm{w}%
_{lm\sigma}^{+}\left(  \bm{R}\right)  \delta_{a}\left(  \bm{R}\right)
\right\vert \bm{G}_{0}\left\vert \bm{w}_{l^{\prime}m^{\prime}%
\sigma^{\prime}}^{+}\left(  \bm{R}^{\prime}\right)  \delta_{a}\left(
\bm{R}^{\prime}\right)  \right\rangle . \label{def Gdd}%
\end{equation}
Its matrix elements may be found in the references \cite{Cichocki2002three},
\cite{felderhof1989displacement}.
It is worth mentioning that
$\left[  G_{0}\left(  \bm{R},\bm{R}^{\prime}\right)  \right]
_{lm\sigma,l^{\prime}m^{\prime}\sigma^{\prime}}$
depends on the difference of positions
$\bm{R}-\bm{R}^{\prime}$
and for nonoverlapping configurations,
i.e. $\left| \bm{R}-\bm{R}^{\prime} \right| \geq 2a$,
it scales as $1/\left|\bm{R}-\bm{R}^{\prime} \right|^{l+l'+\sigma+\sigma'-1}$.
To obtain equations
(\ref{upr rozwiazania mult}) we also used the following definition
\cite{felderhof1988sedimentation}%
\begin{equation}
\hat{Z}_{0}\left(  i\right)  =Z_{0}\left(  i\right)  -Z_{0}\left(  i\right)
\mu_{0}\left(  i\right)  Z_{0}\left(  i\right)  .
\end{equation}

In the multipole formalism the integral equations (\ref{upr rozwiazania mult}%
)\ may be easily reformulated into equation%
\begin{equation}
s_{i}=M\left(  \bm{R}_{i}\right)  \left(  \psi_{0}\left(  \bm{R}%
_{i}\right)  +\sum_{j\not =i}G\left(  \bm{R}_{i},\bm{R}_{j}\right)
s_{j}\right)  \label{s multipole uklad}%
\end{equation}
in the same way, as equations (\ref{ui mobility problem}) and
(\ref{Fi mobility problem}) were transformed into the equation (\ref{s uklad}%
). Moreover definitions of $s_{i}$, $\psi_{0}$ and $M$ are similar to
definitions (\ref{def S}), (\ref{def psi0}) and (\ref{def M}):%
\begin{equation}
s_{i}=%
\begin{bmatrix}
U_{i}\\
F_{i}%
\end{bmatrix}
,
\end{equation}%
\begin{equation}
\psi_{0}\left(  \bm{R}\right)  =%
\begin{bmatrix}
f_{ext}\left(  \bm{R}\right) \\
v_{0}\left(  \bm{R}\right)
\end{bmatrix}
,
\end{equation}%
\begin{equation}
M\left(  i\right)  =%
\begin{bmatrix}
\mu_{0}\left(  i\right)  & \mu_{0}\left(  i\right)  Z_{0}\left(  i\right) \\
Z_{0}\left(  i\right)  \mu_{0}\left(  i\right)  & -\hat{Z}_{0}\left(
i\right)
\end{bmatrix}
.
\end{equation}
and the Green function $G$ in extended multipole space has the following form:%
\begin{equation}
G=%
\begin{bmatrix}
0 & 0\\
0 & G_{0}%
\end{bmatrix}
.
\end{equation}

\section*{References}
\bibliographystyle{unsrt}

\end{document}